\documentclass[iop,apj]{emulateapj} 
\usepackage[dvipsnames]{xcolor}  
\usepackage{amsmath}  
\usepackage{graphicx}
\usepackage{natbib}  
\usepackage[colorlinks=true,
            breaklinks=true,
            linkcolor=red,
            citecolor=MidnightBlue,
            urlcolor=MidnightBlue]{hyperref}
\usepackage[T1]{fontenc}  

\newcommand{\boldface}[1]{#1}

\newcommand\phcms{photon~cm$^{-2}$~s$^{-1}$}
\newcommand\ergs{erg~s$^{-1}$}
\newcommand\ergcms{erg~cm$^{-2}$~s$^{-1}$}

\newcommand{\solarmass}{$M_{\sun}$}

\newcommand{\mbh}{$M_{\rm BH}$}
\newcommand{\msigma}{$M_{\rm BH}-\sigma$}
\newcommand{\ssigma}{$\sigma_{\ast}$}
\newcommand{\eddr}{$\lambda_{\rm Edd}$}

\newcommand{\nhh}{$N_{\rm H,HR}$}
\newcommand{\nhgal}{$N_{\rm H,0}$}
\newcommand{\nhint}{$N_{\rm H}$}

\newcommand{\Fx}{$F_{\rm X}$}
\newcommand{\lx}{$L_{\rm X}$}
\newcommand{\lha}{$L_{\rm H\alpha}$}
\newcommand{\lbol}{$L_{\rm bol}$}
\newcommand{\ledd}{$L_{\rm Edd}$}

\newcommand{\hbeta}{H{$\beta$}}

\newcommand{\halpha}{H{$\alpha$}}

\newcommand{\OIIIb}{[O\,{\sc iii}]\,$\lambda$5007}

\newcommand{\NeV}{[Ne\,{\sc v}]}

\newcommand{\OI}{[O\,{\sc i}]}
\newcommand{\OIwave}{[O\,{\sc i}]\,$\lambda$6300}

\newcommand{\NIIwave}{[N\,{\sc ii}]\,$\lambda$6583}

\newcommand{\SIIwave}{[S\,{\sc ii}]\,$\lambda\lambda$6716,\,6731}

\newcommand{\HII}{H\,{\sc ii}}

\newcommand{\GalNum}{719}        
\newcommand{\AGNNum}{314}           
\newcommand{\onesec}{228}          
\newcommand{\AGNFrac}{43.7\%}       

\newcommand{\SigmaNum}{419}         
\newcommand{\SigmaHoNum}{249}       
\newcommand{\SigmaHyperNum}{163}    

\newcommand{\ClassNum}{418}         
\newcommand{\ClassHoNum}{249}       
\newcommand{\ClassVeronNum}{44}     
\newcommand{\ClassAGN}{243}         

\newcommand{\SeyNum}{59}
\newcommand{\LinNum}{66}
\newcommand{\TraNum}{41}
\newcommand{\HiiNum}{163}
\newcommand{\AbpNum}{89}

\newcommand{\TraAGN}{28}
\newcommand{\HiiAGN}{51}
\newcommand{\AbpAGN}{55}

\newcommand{\SpecNum}{156}       
\newcommand{\WPLNum}{84}            
\newcommand{\PLMENum}{65}             
\newcommand{\pcfNum}{22}            
\newcommand{\IronNum}{27}            
\newcommand{\NHNum}{91}            

\begin{document}
\title{Chandra Survey of Nearby Galaxies: The Catalog}

\author{Rui She\altaffilmark{1},
Luis C.\ Ho\altaffilmark{2,3}, and
Hua Feng\altaffilmark{1}}

\altaffiltext{1}{Department of Engineering Physics and Center for Astrophysics, Tsinghua University, Beijing 100084, China}
\altaffiltext{2}{Kavli Institute for Astronomy and Astrophysics, Peking University, Beijing 100087, China}
\altaffiltext{3}{Department of Astronomy, Peking University, Beijing 100087, China}

\tablefontsize{\scriptsize}

\begin{abstract}

We searched in the public archive of the {\it Chandra} X-ray Observatory as of March 2016 and assembled a sample of  \GalNum\ galaxies within 50 Mpc with ACIS observations available. By cross-correlation with the optical or near-infrared nuclei of these galaxies, \AGNNum\ of them are identified to have an X-ray active galactic nucleus (AGN). The majority of them are low-luminosity AGNs and are unlikely X-ray binaries based upon their spatial distribution and luminosity functions. The AGN fraction is around 60\% for elliptical galaxies and early-type spirals, but drops to roughly 20\% for Sc and later types, consistent with previous findings in the optical.  However, the X-ray survey is more powerful in finding weak AGNs, especially from regions with active star formation that may mask the optical AGN signature. For example, 31\% of the \HII\ nuclei are found to harbor an X-ray AGN. For most objects, a single power-law model subject to interstellar absorption is adequate to fit the spectrum, and the typical photon index is found to be around 1.8. For galaxies with a non-detection, their stacked {\it Chandra} image shows an X-ray excess with a luminosity of a few times $10^{37}$~\ergs\ on average around the nuclear region, possibly composed of faint X-ray binaries. This paper reports on the technique and results of the survey; in-depth analysis and discussion of the results will be reported in forthcoming papers. 

\end{abstract}


\section{Introduction}
\label{sec:intro}

It is widely accepted that every galaxy with a bulge harbors a central supermassive black hole \citep[for a review see][]{kormendy13}. Correlations between the mass of the central black hole ($M_{\rm BH}$) and the properties of the bulge have been identified. For example, $M_{\rm BH}$ scales with the stellar mass or luminosity of the bulge \citep{kormendy95,magorrian98}, and with its stellar velocity dispersion \citep[$\sigma$;][]{gebhardt00,ferrarese00}. Among elliptical galaxies and classical bulges, both correlations are equally tight \citep{kormendy13}.  These correlations lead to the hypothesis that massive galaxies may have evolved along with their central black holes, perhaps linked by feedback from active galactic nuclei \citep[AGNs; for reviews, see][]{alexander12,kormendy13,heckman14}. However, whether the \msigma\ relation extends to lower mass black holes and galaxies is unclear.  Compared with ellipticals and classical bulges, pseudobulges exhibit a significantly different $M_{\rm BH} - \sigma$ relation, with larger scatter and lower normalization \citep{kormendy13,greene16}. The situation for bulgeless galaxies is even more uncertain because the fraction of them that contains a central black hole is still undetermined. The bulgeless, late-type spiral NGC 4395 hosts a black hole with $M_{\rm BH} \approx 10^5$ $M_\sun$  \citep{filippenko03,peterson05,denbrok15}, as does the dwarf spheroidal galaxy POX 52 \citep{barth04}.  A systematic search of the Sloan Digital Sky Survey uncovered 200$-$300 new cases of low-mass ($M_{\rm BH} < 10^6$ $M_\sun$) black holes residing in low-mass galaxies \citep{greene04,greene07b,dong12}, but the statistical completeness of these optical searches are difficult to quantify \citep{greene09}.  As the case of M~33 illustrates \citep{gebhardt01, merritt01}, clearly not every late-type galaxy contains a central black hole.

The fraction of black holes in low-mass galaxies in the local universe can constrain the mechanism for the formation of the supermassive black holes in the early universe \citep{volonteri08}. Specifically, a higher black hole occupation fraction in low-mass ($M_\ast < 10^{10} \; M_\sun$) galaxies is expected if seed black holes were created through the collapse of Population III stars \citep{heger03} instead of direct collapse of gas clouds \citep{haehnelt93}. Based on a spectroscopic survey of a complete sample of nearby galaxies in the northern hemisphere, \citet{ho97,ho97b} report that the fraction of optical AGNs is as high as 60\% for galaxies with a bulge component (Hubble types earlier than Sc),  but drops to less than 20\% for late-type (Sc or later) galaxies.  AGN identification in the optical band may suffer from contamination by star formation, dust obscuration, and general dilution by host galaxy starlight.  The problem is especially severe for galaxies with low-luminosity AGNs (LLAGNs).  

The shortcomings of optical AGN searches can be alleviated, at least in part, by observations at other wavelengths.  High-ionization lines (e.g., \NeV\ $\lambda$14.3 $\mu$m) revealed through mid-infrared spectroscopy have successfully uncovered some weak AGNs in late-type galaxies previously missed in the optical \citep{satyapal08,dudik09,goulding09}. The sample sizes are, however, quite limited; moreover, the reliance on the detection of high-ionization lines automatically precludes the possibility of identifying low-ionization sources associated with low accretion rates \citep{ho08,ho09}.  Radio cores can also effectively reveal weak nuclear activity, but the requisite observations at sub-mJy level sensitivity and sub-arcsecond resolution are not always available for large, homogeneous samples of galaxies  \citep{filho06}. 

X-ray observations, particularly at relatively hard energies ($\ge$ 2 keV), provide an effective, robust means of identifying accretion-powered sources across a wide range of galaxy environments and level of nuclear activity.  As stellar processes and hot gas also generate X-rays, even X-ray searches for AGNs are not immune from contamination by host galaxy emission.  LLAGNs pose the greatest challenge.  At sufficiently low accretion rates, even individual X-ray binaries can rival the signal from  the weakest LLAGNs.  However, host galaxy contamination can be minimized by taking advantage of the low background noise and superior angular resolution of the Advanced CCD Imaging Spectrometer (ACIS) onboard {\it Chandra} \citep{weisskopf02}, which has a point-spread function (PSF) with FWHM $\approx 0\farcs5$.  ACIS is extraordinarily sensitive to point sources, even with relatively short exposures (few ks), and its sharp PSF enables us to pinpoint their possible association with nuclear activity in nearby galaxies if the absolute astrometry of the nucleus of the galaxy is well constrained.  This strategy was successfully implemented by \citet{ho01}, who conducted a snapshot survey of 24 nearby galaxies using {\it Chandra}/ACIS, and it has since been adopted in other studies aimed at obtaining a census of weak AGNs (e.g., \citealt{terashima03,dudik05,gallo08,miller12}).

The X-ray search technique has been particularly effective in identifying candidate LLAGNs, presumably low-mass black holes,  in very late-type, low-mass galaxies.   In a {\it Chandra} archival study of 64 nearby galaxies, \citet{desroches09a} found that  20\%$-$25\% of late-type spiral galaxies contain an X-ray core plausibly due to nonstellar activity.  \citet{zhang09} assembled a larger sample consisting of 187 nearby galaxies observed with {\it Chandra} and detected 86 X-ray AGNs, giving a fraction of $\sim$46\%. They also found a dramatic drop of the AGN fraction, from $\sim$60\% in early-type galaxies to $\sim$30\% in late-type galaxies, but they did not address the association of AGNs with bulgeless galaxies.  \citet{greene12} presented preliminary results on the black hole occupation rate in low-mass galaxies using X-ray AGNs in nearby galaxies identified by \citet{desroches09a} and \citet{gallo10}.  \citet{miller15}, based on X-ray observations of $\sim$200 optically selected early-type galaxies, obtained a lower limit of 20\% for the occupation fraction for early-type galaxies ($< 10^{10} M_\sun$). 

This work conducts the largest, most comprehensive X-ray survey to date of nuclear black holes in a sample of 719 galaxies within 50~Mpc, using data from the 
{\it Chandra} archive. Our primary goal is to quantify the incidence of AGN activity as a function of Hubble type, with special emphasis on low-mass, late-type galaxies.  Our sample is $\sim$3 times larger than the largest previous one of its kind. In addition to searching for low-mass black holes in late-type galaxies, this survey will yield fruitful scientific results in many other aspects. For example, the study can help constrain the accretion models of LLAGNs \citep[for a review see][]{yuan14a} and test whether the presence of a bar is essential for triggering AGN activity \citep{kormendy04}.  We will also investigate non-nuclear X-ray sources, in particular ultraluminous X-ray sources \citep{feng11}.  Our study will appear as a series of several papers, each focusing on a specific topic. This paper will discuss the sample construction and spectral classification (\S~\ref{sec:sample}), X-ray data reduction (\S~\ref{sec:detection}), statistical properties (\S~\ref{sec:analysis}), spectral properties (\S~\ref{sec:properties}), and contamination by X-ray binaries (\S~\ref{sec:contamination}).  

\section{Sample Assembly}
\label{sec:sample}

The sample was assembled based on {\it Chandra}/ACIS observations that were publicly available as of March 2016.  We first generated a full list of ACIS observations, and then searched in the NASA/IPAC Extragalactic Database\footnote{http://ned.ipac.caltech.edu} (NED) for galaxies within 50 Mpc whose nuclear positions were less than 8\arcmin\ from the aim point of any {\it Chandra}\ observation. The adopted distances were taken from NED, in the following order of priority: surface brightness fluctuations, Cepheids variables, tip of the red giant branch, Type Ia supernovae, the fundamental plane, Faber-Jackson relation, Tully-Fisher relation, Tully estimate, and D-Sigma relation \citep[for a review see][]{jacoby92}. If more than one reference is available for the distance by the same means, the latest one is selected, unless otherwise specified. 

Whenever possible, we obtain positions of the galaxy nuclei based on measurements from near-infrared images, which suffer from less obscuration by dust or confusion from young star-forming regions.  Most of the data come from the Two-Micron All Sky Survey (2MASS) \boldface{extended source} catalog \citep{skrutskie06}, or NED otherwise.  In a few cases, the NED positions come from radio observations\footnote{Circinus, M~51a, M~58, NGC~1156, NGC~3079, NGC~5128, NGC~5353}.  We discarded galaxies whose nuclear positions in NED were obtained from X-ray observations. 
The task \textit{dmcoords} in CIAO was used to check whether every galaxy was in the {\it Chandra}\ field of view, followed by visual confirmation. 

This results in a sample of \GalNum\ galaxies with 1559 ACIS observations, including 196 objects having multiple observations. The longest exposure was selected for galaxies with multiple observations.\footnote{\boldface{By coadding all available observations for each galaxy, we find that, for 90\% of the galaxies without an AGN detection, the improvement in sensitivity is less than a factor of 3 compared to the longest exposure. Thus, coadding exposures will not improve the detection rate significantly and is not adopted for simplicity.}} We checked for the CCD pileup effect following the {\it Chandra} ABC Guide to Pileup\footnote{http://cxc.harvard.edu/ciao/download/doc/pileup\_abc.pdf}. For M~58, M~87, and M~106, we chose observations (ObsID 406, 1808, and 2340, respectively) with a shorter frame time (0.4~s), in which the pileup effect was reduced to a negligible level. For many others\footnote{Circinus, M~77, M~81, NGC~1052, 2992, 3998, 4051, 4151, 4203, 5128, 5506, 6300, 7172, and NGC~7314.},  the pileup effect is significant even if observations with a shorter frame time or a larger off-axis angle are available. We ran ray-tracing simulations for {\it Chandra} using MARX \citep{davis12} and found that the pileup at the observed level did not lead to a bias in the source positioning.  Thus, for these objects we used the ACIS image only for astrometric measurements; but the X-ray spectral properties of these sources were derived from published {\it XMM-Newton}\ or {\it ASCA }\ data. All of these targets are bright and well-studied AGNs and do not affect the main result of this study, which focuses on weak or previously unknown AGNs.

\begin{figure}[t]
\center
\includegraphics[width=0.7\columnwidth]{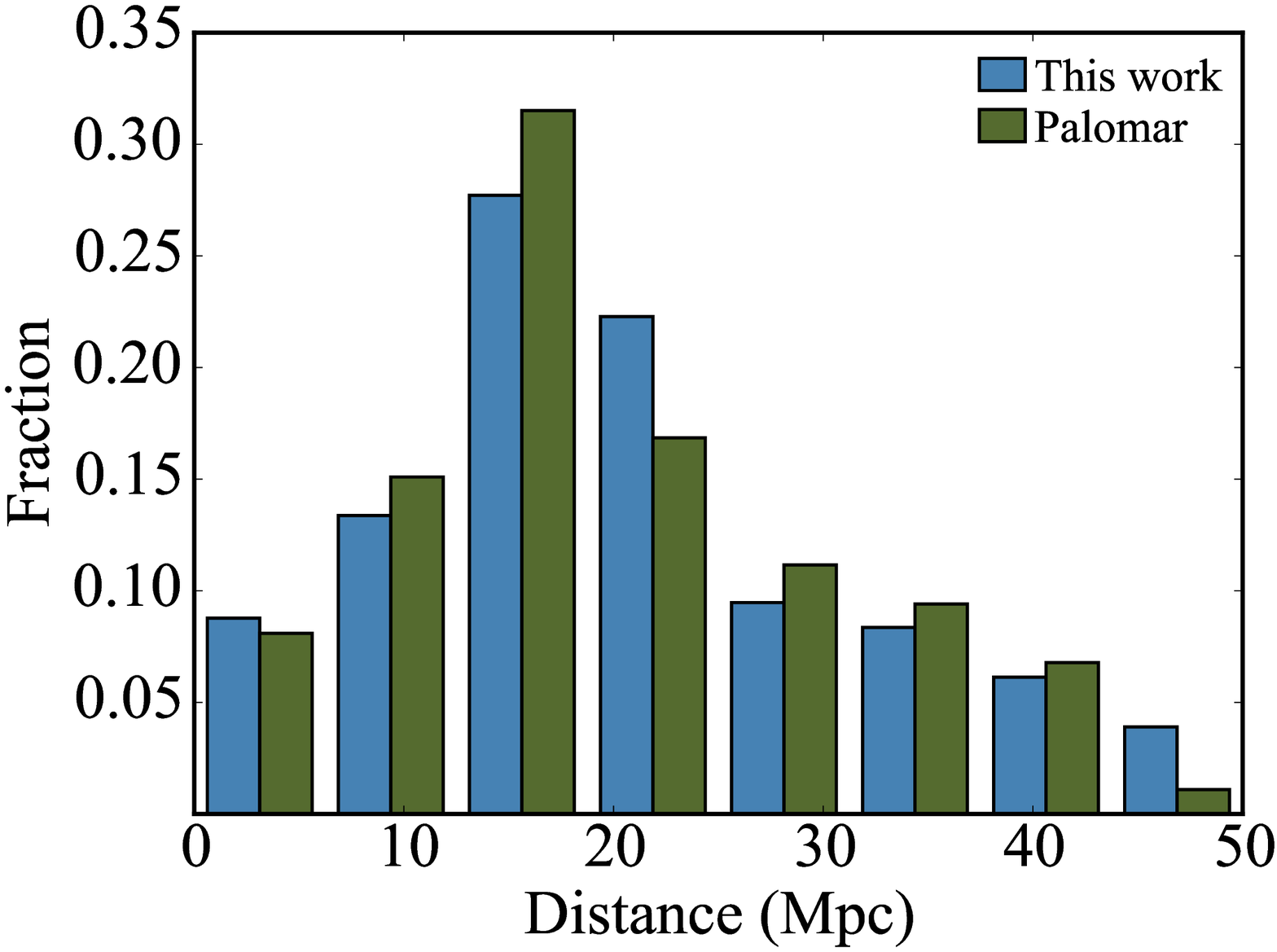}\\
\includegraphics[width=0.7\columnwidth]{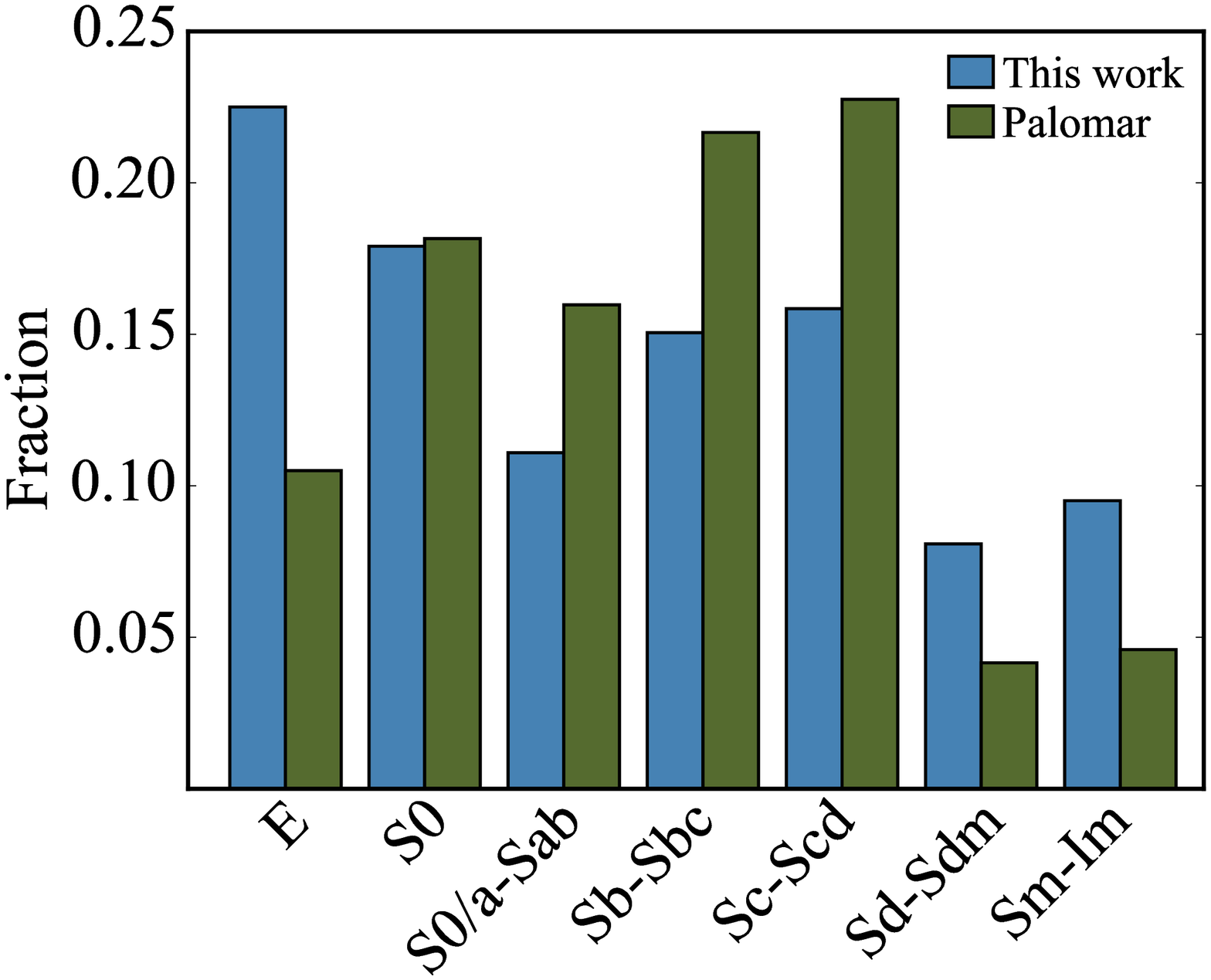}
\caption{Distribution of the distances (top) and Hubble types (bottom) for objects in our sample (blue) and the Palomar survey (green). 
\label{fig:general_dist}}
\end{figure}

\begin{figure*}[bt]
\center
	\includegraphics[width=0.8\textwidth]{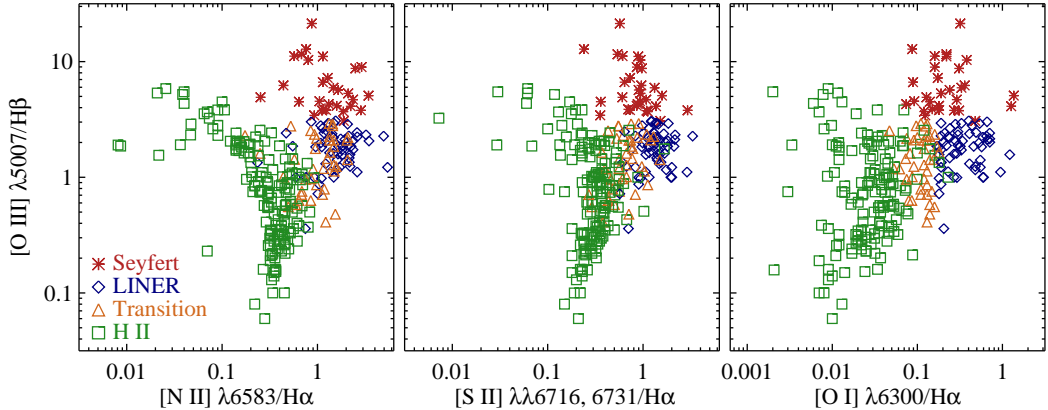}
  \caption{Emission-line diagnostic diagrams for nuclear optical spectral classifications for objects in our sample.
  \label{fig:diagnostic}}
\end{figure*}

The general properties of the galaxies in our sample are listed in Table~\ref{tab:optical_properties}. The Hubble types were adopted from NED, which, in turn, are largely taken from the Third Reference Catalogue of Bright Galaxies (RC3; \citealt{devaucouleurs91}). The \halpha\ luminosities were re-calculated using the \halpha\ fluxes from the Palomar spectroscopic survey of bright galaxies \citep{ho97,ho03} and the distances given in Table~\ref{tab:optical_properties}. There are \SigmaNum\ galaxies in our sample that have published central stellar velocity dispersions $\sigma_\ast$ and their associated uncertainties: \SigmaHoNum\ objects from the Palomar survey \citep{ho09b}, \SigmaHyperNum\ objects from the HyperLeda database\footnote{http://leda.univ-lyon1.fr}, and 7 objects from \citet{gu06}. 

The distributions of the distances and Hubble types for objects in our sample are compared to those in the Palomar survey \citep{ho97b} in Figure~\ref{fig:general_dist}. The Palomar survey is a complete sample of galaxies in the northern hemisphere brighter than 12.5 mag in the $B$ band. The distances for objects in our sample do not show an apparent deviation from those in the Palomar sample. Compared with the Palomar survey, our sample contains a higher fraction of ellipticals and irregular galaxies and has a slightly lower abundance of early-type spirals.

In our sample, \ClassHoNum\ galaxies have nuclear optical spectral classifications available in the Palomar survey \citep{ho97}.  Nuclear spectral classifications for another \ClassVeronNum\  galaxies were found in the catalog of \citet{veron-cetty10}. For galaxies whose spectral classifications are not included in these two catalogs, we adopted nuclear emission-line fluxes from \citet{moustakas06} for 22 objects. In addition, we collected relatively high-quality optical spectra for 103 objects from the literature \citep{kennicutt92,falco99,colless03,jones09,rosales-ortega10,driver11} and performed spectral fitting to determine their spectral classification.  The spectra were fitted using GANDALF\footnote{GANDALF was developed by the SAURON team and is available from the SAURON website (www.strw.leidenuniv.nl/sauron); see \citet{sarzi06} for details.} with Gaussians for the emission lines and a continuum component composed of optimally combined stellar templates of various ages \citep{bruzual03}.  The nuclear spectral classifications were then performed using the diagnostic diagrams defined in \citet{ho97} based on emission-line intensity ratios (\OIIIb/\hbeta\ versus \NIIwave/\halpha, \SIIwave/\halpha, and \OIwave/\halpha)\footnote{In cases when the three diagrams give inconsistent results, the same classification given by two of them was chosen; if none is the same or the other two give ambiguous results, the classification based on \OI\ was chosen.}, shown in Figure~\ref{fig:diagnostic}. (We note that most of the newly analyzed optical spectra are not flux calibrated, so that absolute fluxes of the emission lines are not available.)  To summarize:  \ClassNum\ out of \GalNum\ galaxies in our sample have a nuclear optical spectral classification \citep[for a review see][]{ho08}, including \SeyNum\ Seyferts, \LinNum\ low-ionization nuclear emission-line regions \citep[LINERs;][]{heckman80a}, \TraNum\ transition objects (emission-line nuclei with \OI\ strengths intermediate between those of \HII\ nuclei and LINERs; see \citealt{ho93}), \HiiNum\ \HII\ nuclei, and \AbpNum\ absorption-line nuclei (\boldface{those without optical emission lines}).

\begin{deluxetable*}{lllllllllll}
\tablewidth{\textwidth} \renewcommand{\arraystretch}{1.2}
\tablecaption{general information for galaxies in our 
                sample \label{tab:optical_properties}}
\tablehead{\colhead{ID} & \colhead{Name} & \colhead{Distance} & \colhead{R.A.} & \colhead{Dec.} & \colhead{Class} & \colhead{Hubble Type} & \colhead{\ssigma} & \colhead{$\log$ \lha} & \colhead{$\log$ \mbh} & \colhead{\nhgal}\\ \colhead{ } & \colhead{ } & \colhead{(Mpc)} & \colhead{(J2000)} & \colhead{(J2000)} & \colhead{ } & \colhead{ } & \colhead{(km s$^{-1}$)} & \colhead{(\ergs)} & \colhead{(\solarmass)} & \colhead{($10^{22}$ cm$^{-2}$)}\\ \colhead{(1)} & \colhead{(2)} & \colhead{(3)} & \colhead{(4)} & \colhead{(5)} & \colhead{(6)} & \colhead{(7)} & \colhead{(8)} & \colhead{(9)} & \colhead{(10)} & \colhead{(11)} }
\startdata
1 & AM 0337-353 & 18.9 & 03:39:13.30 & $-$35:22:17.2 & \nodata & $\rm{dSAB0^0?(s)}$ & \nodata & \nodata & \nodata & 0.0158 \\
2 & AM 1247-410 & 36.6 & 12:50:07.74 & $-$41:23:52.8 & \nodata & \nodata & \nodata & \nodata & \nodata & 0.0848 \\
3 & AM 1318-444 & 3.96 & 13:21:47.40 & $-$45:03:42.0 & \nodata & $\rm{IBm}$ & \nodata & \nodata & \nodata & 0.0705 \\
4 & ARK 65 & 10.8 & 01:56:12.01 & +05:35:18.9 & \nodata & \nodata & \nodata & \nodata & \nodata & 0.0435 \\
5 & ARP 244 & 21.5 & 12:01:53.17 & $-$18:52:37.9 & \nodata & \nodata & \nodata & \nodata & \nodata & 0.0315 \\
6 & ARP 261 NED01 & 28.8 & 14:49:30.58 & $-$10:10:23.9 & \nodata & $\rm{IB(s)m\ pec}$ & \nodata & \nodata & \nodata & 0.0785 \\
7 & CCC 61 & 44.2 & 12:48:39.69 & $-$41:16:05.6 & \nodata & \nodata & \nodata & \nodata & \nodata & 0.0854 \\
8 & CCC 111 & 37.7 & 12:49:40.14 & $-$41:21:58.3 & \nodata & \nodata & \nodata & \nodata & \nodata & 0.0848 \\
9 & CCC 123 & 37.3 & 12:49:56.00 & $-$41:24:04.4 & \nodata & \nodata & \nodata & \nodata & \nodata & 0.0848 \\
10 & Circinus Galaxy & 4.21 & 14:13:09.95 & $-$65:20:21.2 & \nodata & $\rm{SA(s)b?}$ & \nodata & \nodata & \nodata & 0.5589 \\
11 & DDO 180 & 22.7 & 13:38:10.32 & $-$09:48:06.6 & \nodata & $\rm{SB(s)m}$ & \nodata & \nodata & \nodata & 0.0319 \\
12 & Draco Dwarf & 0.093 & 17:20:12.39 & +57:54:55.3 & \nodata & $\rm{E\ pec}$ & \nodata & \nodata & \nodata & 0.0225 \\
13 & Dwingeloo 1 & 5.3 & 02:56:50.96 & +58:54:38.8 & \nodata & \nodata & \nodata & \nodata & \nodata & 0.7678 \\
14 & ESO 121- G 20 & 6.08 & 06:15:54.19 & $-$57:43:31.6 & \nodata & $\rm{dwarf}$ & \nodata & \nodata & \nodata & 0.0378 \\
15 & ESO 137- G 6 & 41.6 & 16:15:03.86 & $-$60:54:26.1 & \nodata & $\rm{E1}$ & 355.5$\pm$15.3 & \nodata & 9.61$_{-0.10}^{+0.10}$ & 0.1747 \\
16 & ESO 138- G 10 & 14.7 & 16:59:02.95 & $-$60:12:57.7 & \nodata & $\rm{SA(s)dm}$ & \nodata & \nodata & \nodata & 0.1444 \\
17 & ESO 233- G 35 & 47.1 & 20:09:25.61 & $-$48:17:05.2 & \nodata & $\rm{S0}$ & \nodata & \nodata & \nodata & 0.0390 \\
18 & ESO 293-IG 034 & 20.9 & 00:06:19.91 & $-$41:29:59.7 & \nodata & $\rm{SB(s)cd\ pec\ }$ & \nodata & \nodata & \nodata & 0.0132 \\
19 & ESO 322- G 93 & 47.7 & 12:49:04.15 & $-$41:20:20.4 & \nodata & $\rm{S?}$ & \nodata & \nodata & \nodata & 0.0854 \\
20 & ESO 322- G 102 & 30.5 & 12:49:37.84 & $-$41:23:17.6 & \nodata & $\rm{SB0^0(s)?\ }$ & 104.6$\pm$3.7 & \nodata & 6.85$_{-0.08}^{+0.08}$ & 0.0848
\enddata
\tablecomments{
        Column 1: Source ID. 
        Column 2: Galaxy names. 
        Column 3: Distance from NED. 
        Column 4: Right ascension of galaxies in J2000. 
        Column 5: Declination of galaxies in J2000. 
        Column 6: Spectral classification. 
        ``:'' and ``::'' indicate a classification which is uncertain and highly uncertain, respectively, \citet{ho97}.
        Column 7: Hubble type from NED.
        Column 8: Velocity dispersion.
        Column 9: \halpha\ luminosity from the Palomar survey \citep{ho97,ho03}. 
        The typical uncertainties is 10-30\%, 
        and the letters ``b'', ``c'', ``l'' and ``u'' indicate an uncertainty of 30-50\%, 100\%, a 3$\sigma$ lower limit, 
        and a 3$\sigma$ upper limit, respectively.
        Column 10: Black hole masses estimated using the \msigma\ relation. 
        More details are in the text. 
        Column 11: Galactic absorption column density.
        Table \ref{tab:optical_properties} is published in its entirety in the machine-readable format.
        A portion is shown here for guidance regarding its form and content.
        } 
\end{deluxetable*}

\section{X-ray Data Reduction}
\label{sec:detection}

The ACIS data were reduced using CIAO 4.5 with CALDB 4.5.7.  New level 2 event files were generated using the task \textit{chandra\_repro}.  Intervals with background flaring were excluded using the task \textit{lc\_clean}. For most observations, intervals with background fluxes 1.2 times above or below the mean flux were rejected, as recommended.  For those with heavy flares, \textit{lc\_sigma\_clip} was used instead, and intervals where the flux was over $\pm 3\sigma$ fluctuation of the mean were excluded.  The task \textit{fluximage} was used to extract exposure-corrected images and exposure maps over the energy range 0.3$-$8 keV from the cleaned event files.  A power-law spectrum with a photon index $\Gamma=1.7$, typical for AGNs in nearby galaxies \citep{ho08}, subject to Galactic absorption along the line-of-sight \citep{kalberla05} was used as a weight for the exposure map.  The task \textit{mkpsfmap} was used to create a map for the PSF sizes (radii enclosing 39.3\% of the PSF power) at the average energy of the weighting spectrum for each observation. Source detection was performed with \textit{wavdetect}, which generated source regions that enclose 98.9\% (3$\sigma$) of the total counts for spectral extraction. 

We consider a compact X-ray source to be a candidate AGN if its X-ray position is spatially coincident with the near-infrared/optical position of the stellar nucleus of the galaxy at 99\% confidence level. The relative position uncertainty of an X-ray source with respect to its near-infrared/optical counterpart consists of three components: the near-infrared/optical position error, the absolute astrometry of {\it Chandra}, and the statistical error for the X-ray position. For most objects, the position of the galaxy center is adopted from the 2MASS catalog \boldface{for extended sources}. In 2MASS near-infrared images, the stellar nucleus often appears extended, and the 1$\sigma$ position error on the peak is quoted as $0\farcs5$ \citep{jarrett00}.  Assuming a Rayleigh distribution, this can be translated to a 99\% position uncertainty $\Delta_{\rm 2MASS}=3.05\sigma=1\farcs525$ in radius.  The absolute {\it Chandra} astrometry\footnote{http://cxc.harvard.edu/cal/ASPECT/celmon/} for ACIS is $\Delta_{\rm CXC} = 0\farcs8$ on-axis at a confidence level of 99\%. For sources at an off-axis angle greater than 3\arcmin, an additional position error due to PSF blurring and asymmetry is introduced, roughly equal to 1/4 of the 50\% encircled energy radius at the corresponding off-axis angle. The 99\% statistical error radius from source detection given by \textit{wavdetect} is denoted as $\Delta_{\rm STA}$, which is often much smaller than $\Delta_{\rm CXC}$ and $\Delta_{\rm 2MASS}$. Therefore, the total relative position uncertainty between the X-ray and near-infrared/optical positions at a 99\% confidence level can be expressed as $\Delta_{99\%} = \sqrt{\Delta_{\rm CXC}^2+\Delta_{\rm STA}^2+\Delta_{\rm 2MASS}^2}$.

\boldface{We also compared the NED positions with the 2MASS positions and found that 20 galaxies have a deviation larger than 2\arcsec. Most of them are late-type galaxies with NED positions derived from optical images.  As discussed above, we choose to trust the 2MASS positions in these cases, except for two objects discussed below.  The nucleus of NGC~3621 looks like a point-like source on 2MASS images, and the position from the 2MASS extended source catalog may not be reliable. The nuclear position from NED is consistent with that in the 2MASS point source catalog, and coincident with an X-ray source. This object is then classified as an AGN candidate. NGC~891 is a nearly edge-on spiral, and the nucleus is masked by a dust lane, which is visible on 2MASS images even in the $K_{\rm s}$ (2.16 $\micron$) band, challenging the validity of using its position. On {\it Spitzer} images at 3.6 $\micron$, the dust lane is gone, and the centroid of the nucleus is found to be R.A. = 2$^{\rm h}$22$^{\rm m}$33\fs03, Dec. =  $+$42\arcdeg20\arcmin53\farcs2 (J2000.0), which is $0\farcs76$ away from the 2MASS position for extended sources. We therefore quote this as the center for NGC 891.  An X-ray counterpart is found around it and is still classified as an AGN candidate.}

\begin{figure}[tb]
  \center
	\includegraphics[width=0.9\columnwidth]{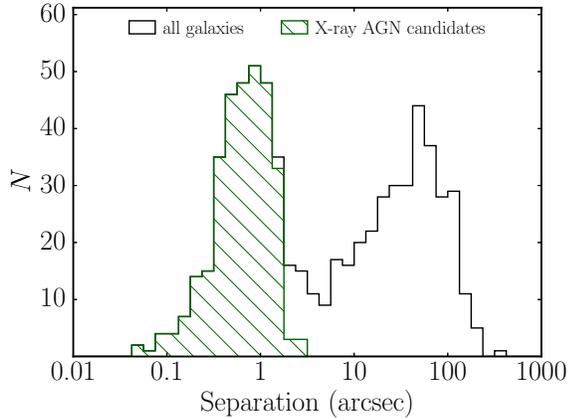}
  \caption{Distribution of the angular distance between the near-infrared or optical nucleus and the nearest point-like X-ray source on the sky plane. The open histogram is for all galaxies in our sample, while the hatched histogram is for AGN candidates, whose X-ray position is coincident with the near-infrared/optical nucleus at 99\% confidence level. 
  \label{fig:offsets}}
\end{figure}

The distribution of the angular separation between the near-infrared or optical nucleus and the nearest point-like X-ray source is shown in Figure~\ref{fig:offsets}.  The distribution is bimodal, with one peak centered near $0\farcs9$ and the other near 50\arcsec.  The distribution for X-ray AGN candidates, whose separation is less than $\Delta_{99\%}$, is also displayed (hatched histogram). The angular separations, less than 1\arcsec\ for most AGN candidates, are listed in Table~\ref{tab:x-ray_properties}. 

\subsection{Count Rates and Hardness Ratios}

Source photons were extracted from the regions defined by \textit{wavdetect}. The background was estimated from a concentric annulus surrounding the source ranging from 5\arcsec\ to 10\arcsec\ for most cases, or from another circular region if there is source confusion in the annulus. The net count rates and their associated errors were estimated assuming Poisson distributions for events in both the source and background regions.  Given the counts, areas, and fractions of PSF powers in both the source and background apertures, the task \textit{aprates} was used to calculate the net count rate and its 90\% error bound with the Bayesian approach. In our cases, the PSF fraction is 0.989 in the source region and assumed to be 0 in the background region. For objects with no AGN identification, a circular source aperture with a radius that encloses 90\% of the energy centered at the near-infrared/optical position of the nucleus was used to estimate the 90\% upper limit of the source count rate using the task \textit{aprates}. 

We calculated the hardness ratios $H_0$, $H_1$ and $H_2$, defined as $H_0 \equiv (C_{\rm H}-C_{\rm M}-C_{\rm S})/C_{\rm T}$, $H_1 \equiv (C_{\rm M}-C_{\rm S})/C_{\rm T}$, and $H_2 \equiv (C_{\rm H}-C_{\rm M})/C_{\rm T}$, where $C_{\rm S}$, $C_{\rm M}$,  $C_{\rm H}$, and $C_{\rm T}$ are counts in the soft band (0.3$-$1~keV), medium band (1$-$2~keV), hard band (2$-$8~keV), and full band (0.3$-$8~keV), respectively. The hardness ratio $H_0$ was used to infer the intrinsic hydrogen absorption column density \nhh\, assuming a fixed photon index of $\Gamma=1.8$, a typical value measured in this work (see \S~\ref{subsec:gamma}). We estimate the uncertainty of \nhh\ by varying $\Gamma$ between 1.3 and 2.3, which encompasses most of the values observed in this work.

\tablefontsize{\scriptsize}
\begin{turnpage}
\begin{deluxetable*}{llrcrccrrrrrrc}
\tablewidth{0pc}  \renewcommand{\arraystretch}{1.2}
\tablecaption{X-ray properties for galaxies in our sample \label{tab:x-ray_properties}}
\tablehead{\colhead{ID} & \colhead{Name} & \colhead{Obsid} & \colhead{Inst} & \colhead{Exposure} & \colhead{Offset} & \colhead{AGN} & \colhead{$H_0$} & \colhead{$H_1$} & \colhead{$H_2$} & \colhead{\nhh} & \colhead{\Fx} & \colhead{$\log$ \lx} & \colhead{Note}\\ \colhead{ } & \colhead{ } & \colhead{ } & \colhead{ } & \colhead{(ks)} & \colhead{($^{\prime\prime}$)} & \colhead{ } & \colhead{ } & \colhead{ } & \colhead{ } & \colhead{($10^{22}$ cm$^{-2}$)} & \colhead{($10^{-14}$~cgs)} & \colhead{(\ergs)} & \colhead{ }\\ \colhead{(1)} & \colhead{(2)} & \colhead{(3)} & \colhead{(4)} & \colhead{(5)} & \colhead{(6)} & \colhead{(7)} & \colhead{(8)} & \colhead{(9)} & \colhead{(10)} & \colhead{(11)} & \colhead{(12)} & \colhead{(13)} & \colhead{(14)}  }
\startdata
1 & AM 0337-353 & 624 & ACIS-S & 43.6 & \nodata & N & \nodata & \nodata & \nodata & \nodata & $<$0.75 & $<$37.52 & G1.8 \\
2 & AM 1247-410 & 8179 & ACIS-S & 29.8 & \nodata & N & \nodata & \nodata & \nodata & \nodata & $<$0.14 & $<$36.49 & G1.8 \\
3 & AM 1318-444 & 15200 & ACIS-I & 19.5 & \nodata & N & \nodata & \nodata & \nodata & \nodata & $<$0.41 & $<$36.60 & G1.8 \\
4 & ARK 65 & 2223 & ACIS-S & 30.4 & \nodata & N & \nodata & \nodata & \nodata & \nodata & $<$0.73 & $<$37.65 & G1.8 \\
5 & ARP 244 & 3041 & ACIS-S & 72.9 & \nodata & N & \nodata & \nodata & \nodata & \nodata & $<$0.04 & $<$35.85 & G1.8 \\
6 & ARP 261 NED01 & 5191 & ACIS-S & 55.0 & \nodata & N & \nodata & \nodata & \nodata & \nodata & $<$0.13 & $<$37.28 & G1.8 \\
7 & CCC 61 & 16223 & ACIS-S & 179.0 & \nodata & N & \nodata & \nodata & \nodata & \nodata & $<$0.12 & $<$38.21 & G1.8 \\
8 & CCC 111 & 8179 & ACIS-S & 29.8 & \nodata & N & \nodata & \nodata & \nodata & \nodata & $<$0.37 & $<$38.52 & G1.8 \\
9 & CCC 123 & 8179 & ACIS-S & 29.8 & \nodata & N & \nodata & \nodata & \nodata & \nodata & $<$0.12 & $<$37.63 & G1.8 \\
10 & Circinus Galaxy & 12823 & ACIS-S & 152.4 & 0.35 & Y & $0.484\pm$0.005 & $0.203\pm$0.006 & $0.511\pm$0.005 & 1.73$_{-0.64}^{+0.68}$ & 1400.00 & 42.20 & Ref.2 \\
11 & DDO 180 & 6323 & ACIS-S & 7.1 & \nodata & N & \nodata & \nodata & \nodata & \nodata & $<$0.51 & $<$38.23 & G1.8 \\
12 & Draco Dwarf & 9568 & ACIS-S & 24.5 & \nodata & N & \nodata & \nodata & \nodata & \nodata & $<$0.20 & $<$32.95 & G1.8 \\
13 & Dwingeloo 1 & 7151 & ACIS-I & 25.3 & \nodata & N & \nodata & \nodata & \nodata & \nodata & $<$0.20 & $<$36.53 & G1.8 \\
14 & ESO 121- G 20 & 14351 & ACIS-I & 26.2 & \nodata & N & \nodata & \nodata & \nodata & \nodata & $<$0.62 & $<$36.70 & G1.8 \\
15 & ESO 137- G 6 & 8178 & ACIS-S & 57.4 & 0.29 & Y & $-0.856\pm$0.009 & $-0.008\pm$0.018 & $-0.388\pm$0.016 & $<$0.01 & 5.24$_{-0.81}^{+1.11}$ & 39.92$_{-0.13}^{+0.10}$ & Fit \\
16 & ESO 138- G 10 & 14800 & ACIS-S & 9.8 & \nodata & N & \nodata & \nodata & \nodata & \nodata & $<$0.26 & $<$37.52 & G1.8 \\
17 & ESO 233- G 35 & 3191 & ACIS-I & 23.5 & \nodata & N & \nodata & \nodata & \nodata & \nodata & $<$2.17 & $<$39.49 & G1.8 \\
18 & ESO 293-IG 034 & 11236 & ACIS-S & 9.9 & 1.29 & Y & $0.130\pm$0.205 & $0.217\pm$0.202 & $0.217\pm$0.202 & 1.11$_{-0.82}^{+1.17}$ & 1.91$_{-0.81}^{+1.10}$ & 39.45$_{-0.38}^{+0.45}$ & HR \\
19 & ESO 322- G 93 & 16223 & ACIS-S & 179.0 & 0.25 & Y & $-0.300\pm$0.122 & $0.500\pm$0.111 & $-0.233\pm$0.125 & $<$0.63 & 0.08$_{-0.07}^{+0.07}$ & 38.46$_{-0.48}^{+0.32}$ & HR \\
20 & ESO 322- G 102 & 8179 & ACIS-S & 29.8 & \nodata & N & \nodata & \nodata & \nodata & \nodata & $<$0.61 & $<$38.62 & G1.8
\enddata
\tablecomments{
        Column 1: Source ID.
        Column 2: Galaxy names.
        Column 3: Chandra observation ID.
        Column 4: Focal plane instrument.
        Column 5: Exposure time.
        Column 6: Offset from the nearest point-like x-ray source to the optical position of the galaxy nucleus.
        Column 7: The galaxy has an AGN or not.
        Columns 8-10: Hardness ratios, defined as $H_0 \equiv (C_{\rm H}-C_{\rm M}-C_{\rm S})/C_{\rm T}$, $H_1 \equiv (C_{\rm M}-C_{\rm S})/C_{\rm T}$, and $H_2 \equiv (C_{\rm H}-C_{\rm M})/C_{\rm T}$, where $C_{\rm S}$, $C_{\rm M}$,  $C_{\rm H}$, and $C_{\rm T}$ are counts in the soft band (0.3$-$1~keV), medium band (1$-$2~keV), hard band (2$-$8~keV), and full band (0.3$-$8~keV), respectively.
        Column 11: Intrinsic absorption inferred from hardness ratio.
        Column 12: Observed flux or 90\% upper limit in the energy band of 2-10 keV.
        Column 13: Log of 2-10 keV luminosity, or 90\% upper limits.
        Column 14: Note for X-ray luminosity. ``Fit'', 
        ``HR'' and ``G1.8'' denote luminosity from spectral fitting,
        hardness ratio, and assumption of $\Gamma$=1.8 with Galactic \nhgal ,
        respectively.  For sources suffering from pileup, the references are given:
        Ref.1: \citet{ebrero11}, Ref.2: \citet{rosa12}, 
        Ref.3: \citet{cappi06}, Ref.4: \citet{shu10}, 
        Ref.5: \citet{iyomoto98}, Ref.6: \citet{ptak04}, 
        Ref.7: \citet{weaver99}, Ref.8: \citet{guainazzi10}, 
        Ref.9: \citet{evans04}. Table \ref{tab:x-ray_properties} is published in its entirety in the machine-readable format.
        A portion is shown here for guidance regarding its form and content.} 
\end{deluxetable*}

\end{turnpage}

\subsection{X-ray Spectra}

X-ray spectra were extracted for AGN candidates that have at least 100 photons in the energy range 0.3$-$8 keV, from the source and background regions defined above.  The task \textit{specextract} was employed to extracted the energy spectra, which were grouped to have at least 15 counts in each spectral bin and fitted in XSPEC 12.8 \citep{arnaud96}. We first tried a simple power-law model subject to interstellar absorption with one absorption component fixed at the Galactic value \citep{kalberla05} and a second free component to account for extragalactic absorption.  If the single power-law model was inadequate to fit the data, resulting in a null hypothesis probability less than 0.05, we tried to add a thermal plasma component ({\tt mekal}) or a cool blackbody component ({\tt bbody}), or we replaced the simple absorption model with a partial absorption model, and then tested the significance of the additional component with the F-test (being significant if the chance probability is $< 0.05$ ).  The two-component model was found to be able to fit most of the spectra adequately. The extragalactic component was removed if the column density for the extragalactic absorption component converged to 0.   

\subsection{Flux, Luminosity, and Eddington Ratio}
\label{subsec:luminosity}

For AGN candidates with sufficient photons for spectral fitting, the observed flux, intrinsic luminosity, and their associated errors in the 2$-$10 keV band were calculated using the \textit{cflux} model in XSPEC.  For those without enough photons for spectral fitting, the flux and luminosity were translated from their source count rate assuming an absorbed power-law model, in which the photon index is fixed at two extremes $\Gamma=1.3-2.3$ (for interval estimate),
the absorption is inferred from the observed hardness ratio $H_0$ (\nhh, see above) and fixed at it, and the local detector response adopted is with the corresponding $\Gamma$. For non-detections in the nuclear region, the 90\% upper limit on the source count rate was converted to an upper limit in flux and luminosity assuming an absorbed power-law spectrum with a photon index $\Gamma=1.8$ and Galactic \nhgal\ based on the local detector responses.

\begin{figure}[tb]
  \center
	\includegraphics[width=0.8\columnwidth]{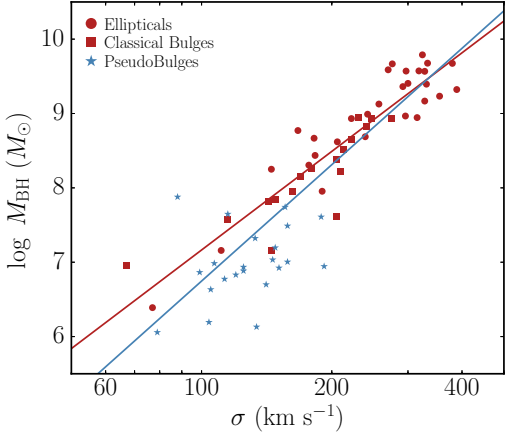}
  \caption{
Black hole masses versus central stellar velocity dispersions \citep[data adopted from][]{kormendy13}. The blue line is the linear regression for all  objects, while the red line is the fit for elliptical galaxies and classical bulges. 
  \label{fig:Msigma}}
\end{figure}

The majority of the objects in our sample do not have a direct measurement of the black hole mass. We thus estimated the mass of black holes using the \msigma\ relation.  As discussed by \citet{kormendy13}, the scaling relations between black hole mass and bulge properties are tight only for classical bulges and elliptical galaxies; pseudobulges show considerably more scatter and an apparently lower zero point.  \citet{kormendy13} provide an updated \msigma\ relation only for classical bulges and ellipticals; they did not explicitly fit the pseudobulges, or their sample as a whole with all objects combined.  We do not have bulge type classifications for the majority of our sample, but we can be certain that it contains both bulge types, as many of our objects have relatively late Hubble types, which preferentially host pseudobulges (e.g., \citealt{kormendy04}).  In order to be able to estimate black hole masses from stellar velocity dispersions, we fit a single \msigma\ relation for all galaxies (independent of bulge type) deemed by \citet{kormendy13} to have a reliable black hole mass measurement, using black hole masses and stellar velocity dispersions as tabulated by those authors.  We performed the linear regression using the ($x|y$) type bivariate correlated errors and intrinsic scatter (BCES) method \citep{akritas96,kelly07}. Errors on both sides and an intrinsic scatter of the relation were taken into account. The regression results in the relation shown in Figure~\ref{fig:Msigma}:
\begin{equation}
\begin{split}
  \log \left( \frac{M_{\rm BH}}{M_{\sun}} \right) = 
  & -(0.68 \pm 0.05) + \\
  & (5.20 \pm 0.37) \log \left( \frac{\sigma}{200\; {\rm km\; s}^{-1}} \right)
  \label{m-sigma}
\end{split}
\end{equation}
with an intrinsic scatter of 0.44~dex.  If we restrict the fit to just ellipticals and galaxies with classical bulges, the relation has a slope of $4.40 \pm 0.32$ and a smaller intrinsic scatter of 0.27~dex (red line in Figure~\ref{fig:Msigma}).  This result is very close to that of \citet{kormendy13} (slope $4.38\pm0.29$ and intrinsic scatter 0.29 dex), who used a different fitting method than ours, verifying that our fitting results are robust with respect to fitting method.  Our global \msigma\ relation (with all galaxies included) is likely to overestimate the black hole masses for pseudobulges and bulgeless galaxies, and therefore underestimate their Eddington ratios; conversely, ellipticals and classical bulges will have their black hole masses underestimated and Eddington ratios overestimated.   While this situation is not ideal, it is the best we can do under the current circumstances.

When applying the \msigma\ relation, we use central stellar velocity dispersions, which are a reasonable proxy of bulge velocity dispersions \citep[see][]{kormendy13}. \boldface{To be consistent with \citet{ho08}, we adopted the empirical bolometric correction for the X-ray band, \lbol = 16\;\lx. \citet{ho08} obtained this estimate based on broadband SEDs of LLAGNs, consistent with the result from \citet{vasudevan07}, who suggested $L_{\rm bol}/L_{\rm X} = 15-25$ for AGNs with \lbol$/$\ledd~$<$~0.1.} The Eddington ratio is then defined as \eddr = 16\;\lx$/$\ledd, where \ledd = $1.26\times 10^{38}$ \mbh$/$\solarmass ~\ergs.

\section{Results and Discussion}
\label{sec:analysis}
Among the \GalNum\ galaxies in our {\it Chandra} sample, \AGNNum\ are identified as AGN candidates (Table~\ref{tab:x-ray_properties}), including \onesec\ that are located on the sky plane less than $1\arcsec$ from the near-infrared/optical nucleus. The total X-ray AGN fraction is \AGNFrac. Among the AGN candidates, optical spectral classifications are available for \ClassAGN\ (77\%) objects,  
\boldface{out of which 44\% (106/\ClassAGN) are \HII\ or absorption-line nuclei that do not show AGN signatures in the optical.} 
Among galaxies previously recognized to contain \HII\ nuclei, 31\% (\HiiAGN/\HiiNum) harbor an X-ray AGN candidate, \boldface{and this fraction is 62\% (\AbpAGN/\AbpNum) for absorption-line nuclei.}  
\boldface{For optical Seyferts and LINERs, 87\% (109/125) of them have an X-ray core.}
The X-ray AGN fraction is as high as $\sim$60\% for early-type (E$-$Sbc) galaxies and drops to $\sim$20\% in late types (Sc and later), consistent with previous results from optical surveys \citep{ho97b,ho08}. \boldface{We summarize the X-ray AGN detection rates in Table~\ref{tab:agnrates}.} A comprehensive analysis of AGN demographics will be given in R. She et al. (in preparation).

\tablefontsize{\footnotesize}
\begin{deluxetable}{lccc}
\tablewidth{\columnwidth}
\tablecolumns{4}
\tablecaption{\boldface{X-ray AGN detection rates}}
\tablehead{
\colhead{Sample}                  &
\colhead{Galaxies}                  &
\colhead{X-ray Cores}                     &
\colhead{Rate}                   }
\startdata
Whole sample                   & \GalNum   & \AGNNum   & 44\% \\
With optical classifications   & \ClassNum & \ClassAGN & 58\% \\
\noalign{\smallskip}\hline\noalign{\smallskip}
Optical Seyferts and LINERs    & 125       & 109       & 87\% \\
Optical transition objects     & \TraNum   & \TraAGN   & 68\% \\
Optical \HII\ nuclei           & \HiiNum   & \HiiAGN   & 31\% \\
Optical absorption-line nuclei & \AbpNum   & \AbpAGN   & 62\%
\enddata
\label{tab:agnrates}
\end{deluxetable}

\begin{figure*}[htbp]
  \center
    \includegraphics[width=2.00\columnwidth]{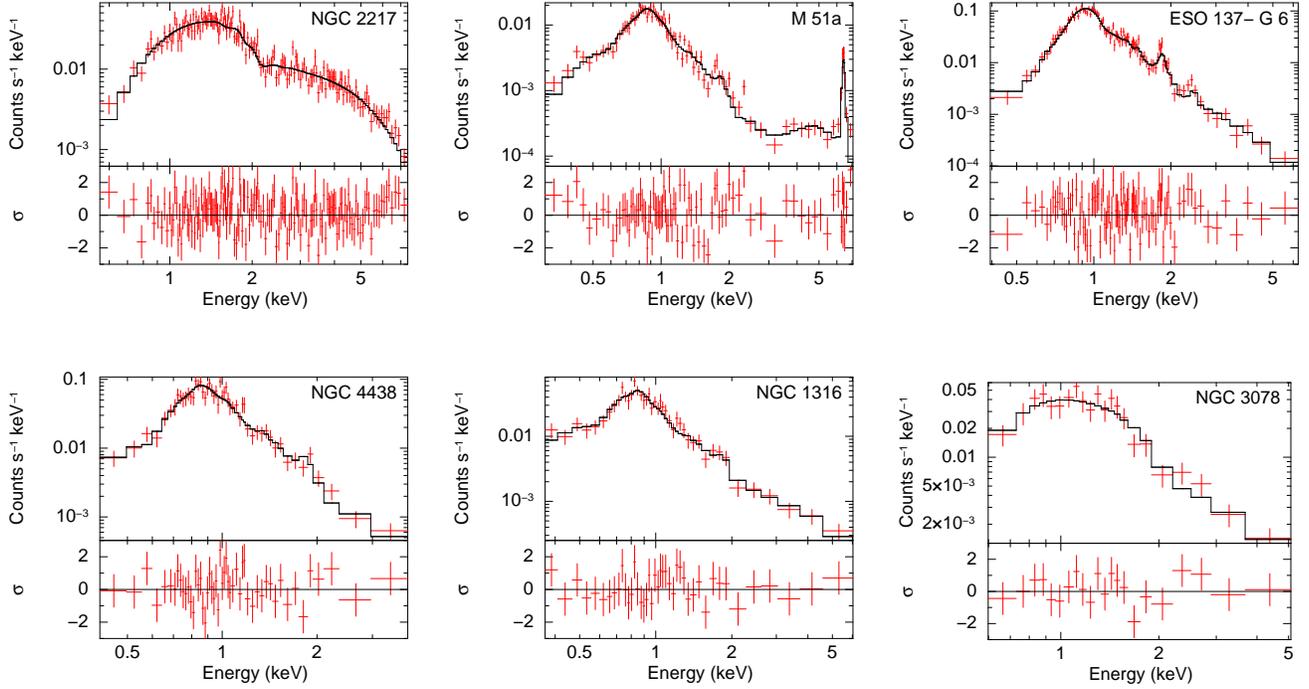}
  \caption{Spectral fits for the individual objects. Figure~\ref{fig:fits} is published in its entirety in electronic format. 6 objects are shown here for representatives.
  \label{fig:fits}}
\end{figure*}

Spectral fitting can be performed for the AGN candidates that have sufficient counts. Among the \SpecNum\ AGN spectra that meet this criterion, \WPLNum\ can be adequately fitted with a single power-law model, and \PLMENum\ with a {\tt powerlaw} plus {\tt mekal} model (a non-solar abundance for {\tt mekal} is preferred in 5 cases). \boldface{A partially covering absorbed power-law model ({\tt pcfabs*powerlaw}) is more favored for \pcfNum\ of them.} An extragalactic absorption component is needed in \NHNum\ of the spectra. \boldface{Other models (such as {\tt pexmon}, {\tt powerlaw+bbody}, and {\tt powerlaw+mekal+bbody}) are used for very few cases.} The best-fit models and parameters are shown in Table~\ref{tab:fit_results}, where the errors are quoted at 90\% confidence level. Figure~\ref{fig:fits} shows the fits for the individual objects.
A statistically significant iron K$\alpha$ line, which we model with a simple Gaussian profile, is detected in \IronNum\ objects (Table~\ref{tab:iron}).

\begin{figure}[htpb]
  \center
    \includegraphics[width=\columnwidth]{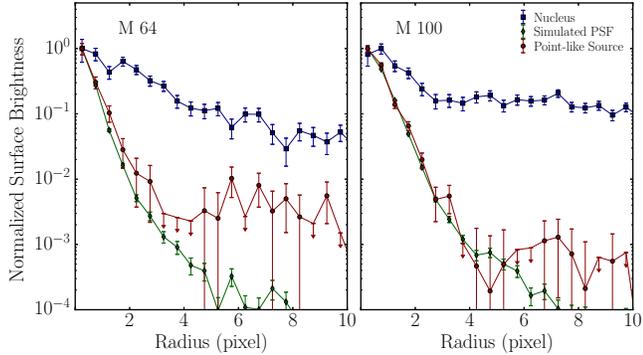}
  \caption{Radial profile for the nuclear X-ray source in M 64 (left) and M 100 (right), compared with a local PSF (simulated with MARX) and a nearby point-like object. They both show a single {\tt mekal} spectrum and appear to be extended sources.
  \label{fig:radialprofile}}
\end{figure}

\begin{figure*}[tb]
\center
\includegraphics[width=0.85\textwidth]{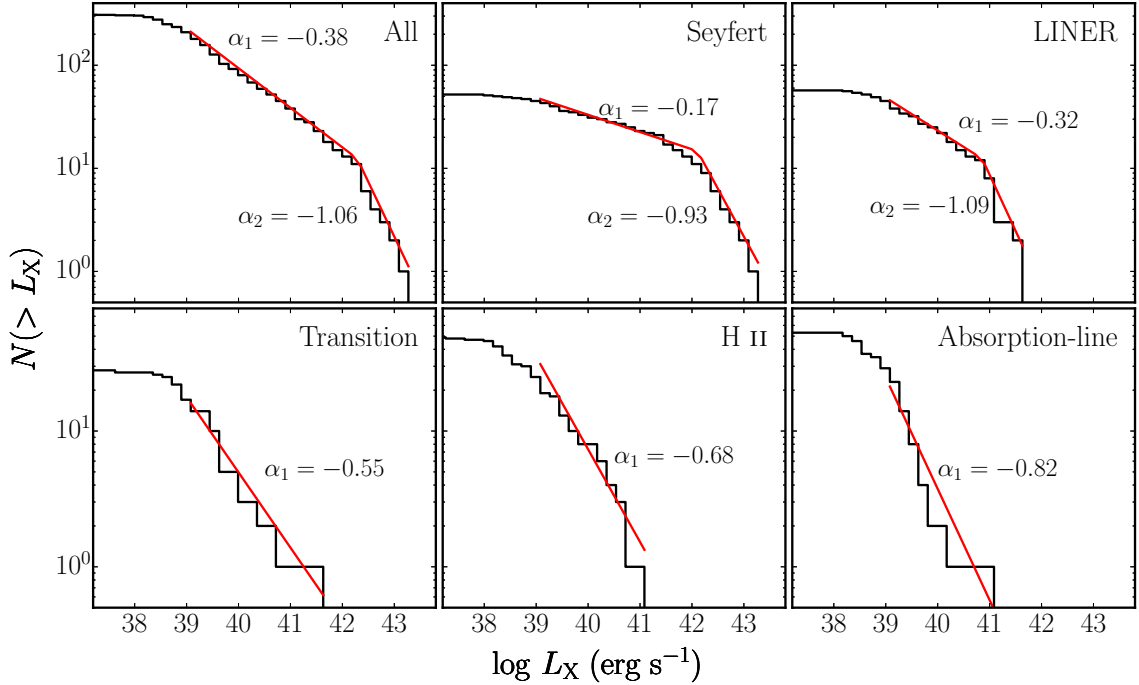}
 \caption{Cumulative luminosity distribution \boldface{(2$-$10 keV band)} for AGN candidates in our sample. The red line is the fit to a broken power-law or a single power-law model at luminosities above $10^{39}$~\ergs.
 \label{fig:lum}}
\end{figure*}

\begin{figure}[tb]
\center
\includegraphics[width=0.8\linewidth]{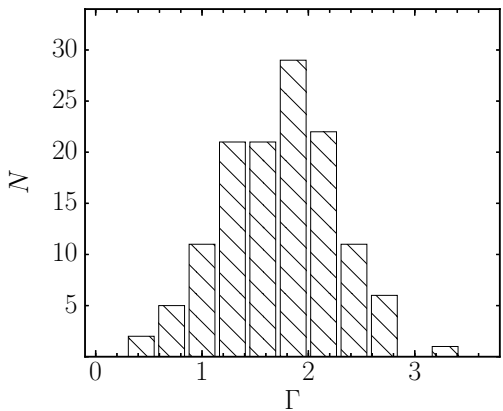}
\caption{Distribution of the power-law photon index $\Gamma$ \boldface{in the 0.3$-$8 keV band}, for AGN candidates whose best-fit model includes a power-law component \boldface{and has a dof $>5$}.
\label{fig:gamma}} 
\end{figure}

\begin{figure}[tb]
\center
\includegraphics[width=0.7\columnwidth]{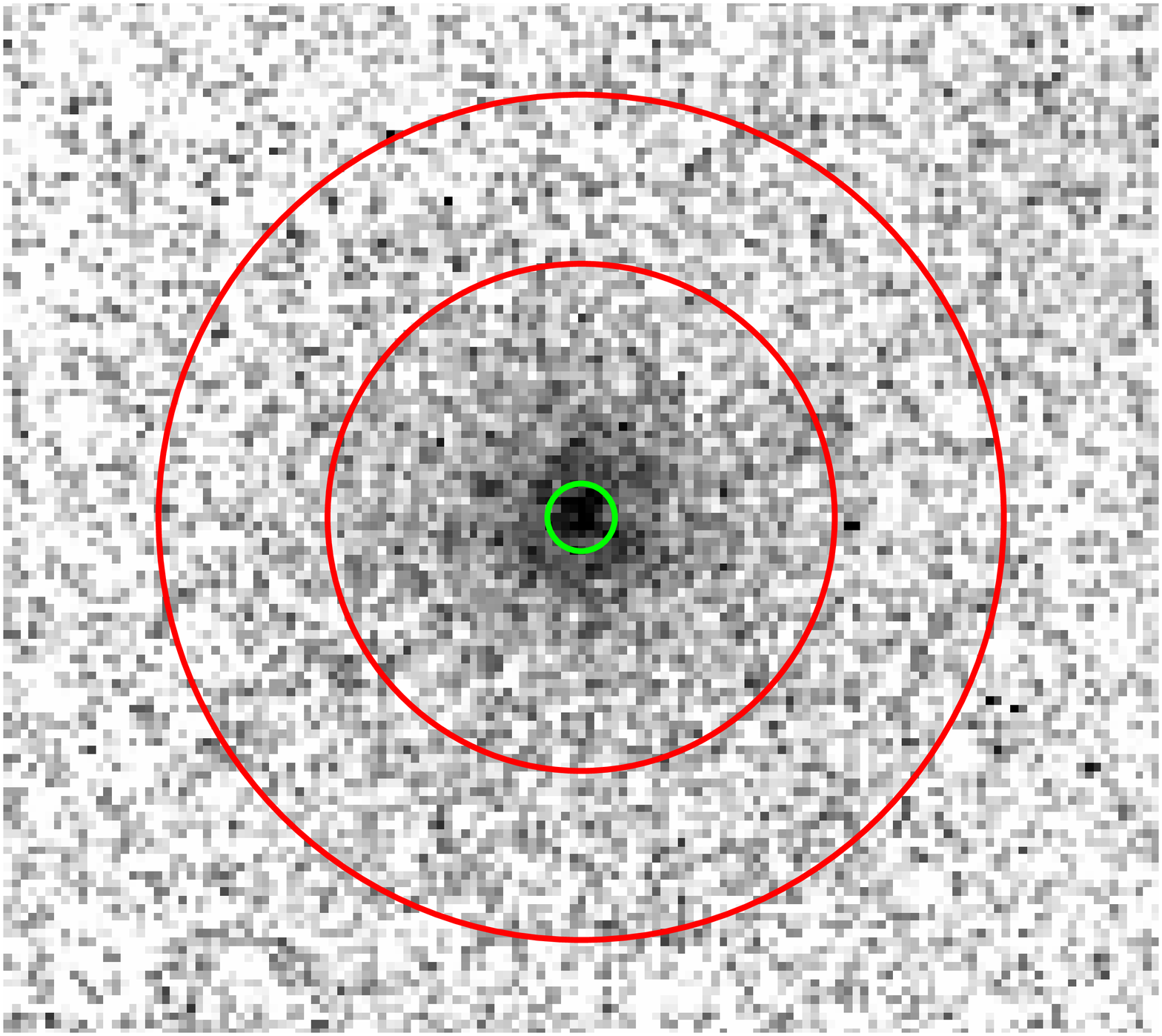}
\caption{Stacked image of 368 non-detections, in the energy range 0.3$-$8 keV. The green circle (2$\arcsec$ radius) is the source aperture, and the red annulus is the background region used to calculate source flux. 
\label{fig:stackimage}}
\end{figure}

%
%

\tablefontsize{\tiny}
\begin{deluxetable*}{lllllllrllrc}
\tablewidth{\textwidth} \renewcommand{\arraystretch}{1.2}
\tablecaption{Fitting results of galaxies \label{tab:fit_results}}
\tablehead{\colhead{ID} & \colhead{Name} & \colhead{Model} & \colhead{$\log$ Norm$_1$} & \colhead{$\log L_1$} & \colhead{$\log$ Norm$_2$} & \colhead{$\log L_2$} & \colhead{\nhint} & \colhead{$\Gamma$} & \colhead{$T$} & \colhead{$\chi^2$/dof} & \colhead{Note}\\ \colhead{ } & \colhead{ } & \colhead{ } & \colhead{ } & \colhead{(\ergs)} & \colhead{ } & \colhead{(\ergs)} & \colhead{($10^{22}$ cm$^{-2}$)} & \colhead{ } & \colhead{(keV)} & \colhead{ } & \colhead{ }\\ \colhead{(1)} & \colhead{(2)} & \colhead{(3)} & \colhead{(4)} & \colhead{(5)} & \colhead{(6)} & \colhead{(7)} & \colhead{(8)} &  \colhead{(9)} & \colhead{(10)} & \colhead{(11)} & \colhead{(12)}}
\startdata
10 & Circinus Galaxy & pileup & \nodata & \nodata & \nodata & \nodata & 430.00$_{-70.00}^{+40.00}$ & 1.56$_{-0.51}^{+0.16}$ & \nodata & \nodata & Ref.2 \\
15 & ESO 137- G 6 & PL+ME & $     -4.52$ & 40.48 & $     -3.84$ & 40.90 & 0.12$_{-0.04}^{+0.04}$ & 2.45$_{-0.34}^{+0.34}$ & 0.75$_{-0.03}^{+0.03}$ & 104.6/88 &  \\
21 & ESO 428- G 14 & PL+ME+BB+G & $     -4.61$ & 40.37 & $     -4.51$ & 39.73 & \nodata & 1.03$_{-0.27}^{+0.26}$ & 0.69$_{-0.07}^{+0.10}$ & 65.7/64 &  \\
24 & ESO 495- G 21 & PL+ME & $     -4.89$ & 39.43 & $     -4.61$ & 38.74 & \nodata & 0.66$_{-0.27}^{+0.28}$ & 0.68$_{-0.05}^{+0.10}$ & 44.8/33 &  \\
40 & IC 342 & PL+VME & $     -4.27$ & 38.64 & $     -3.74$ & 38.68 & \nodata & 1.92$_{-0.25}^{+0.21}$ & 0.66$_{-0.03}^{+0.03}$ & 142.6/106 &  \\
43 & IC 750 & PL & $     -5.17$ & 39.50 & \nodata & \nodata & \nodata & 2.30 & \nodata & 5.2/3 &  \\
47 & IC 1459 & PL & $     -3.64$ & 41.19 & \nodata & \nodata & 0.21$_{-0.02}^{+0.02}$ & 1.86$_{-0.06}^{+0.06}$ & \nodata & 212.7/198 &  \\
51 & IC 2560 & PABS*PL+ME+G & $     -3.25$ & 40.60 & $     -4.74$ & 39.67 & 22.41$_{-4.69}^{+7.28}$ & 2.51$_{-0.26}^{+0.18}$ & 0.19$_{-0.04}^{+0.04}$ & 144.7/73 &  \\
88 & IC 5267 & PL+ME & $     -5.61$ & 39.19 & $     -5.49$ & 38.75 & \nodata & 1.46$_{-0.59}^{+0.55}$ & 0.37$_{-0.06}^{+0.15}$ & 5.6/7 &  \\
100 & MCG -03-34-064 & PABS*PL+ME+G & $     -2.59$ & 42.58 & $     -4.35$ & 40.50 & 38.04$_{-5.97}^{+6.75}$ & 1.89$_{-0.21}^{+0.16}$ & 0.62$_{-0.20}^{+0.09}$ & 63.8/48 &  \\
102 & M 31 & PL+ME & $     -4.14$ & 37.56 & $     -4.77$ & 36.55 & \nodata & 2.66$_{-0.14}^{+0.13}$ & 0.22$_{-0.04}^{+0.05}$ & 81.5/75 &  \\
103 & M 32 & PL & $     -5.02$ & 36.58 & \nodata & \nodata & \nodata & 2.60$_{-0.19}^{+0.20}$ & \nodata & 59.2/32 &  \\
105 & M 49 & PL+ME & $     -5.19$ & 39.10 & $     -4.57$ & 39.44 & \nodata & 2.27$_{-0.18}^{+0.17}$ & 0.73$_{-0.04}^{+0.02}$ & 100.1/80 &  \\
106 & M 51a & PABS*PL+ME+G & $     -3.78$ & 40.00 & $     -5.03$ & 38.33 & 35.16$_{-6.98}^{+8.44}$ & 2.46$_{-0.17}^{+0.15}$ & 0.64$_{-0.03}^{+0.03}$ & 110.3/78 &  \\
107 & M 51b & PL+ME & $     -5.20$ & 38.83 & $     -5.39$ & 37.93 & \nodata & 1.12$_{-0.20}^{+0.21}$ & 0.78$_{-0.19}^{+0.25}$ & 55.0/48 &  \\
108 & M 58 & PL & $     -2.90$ & 41.80 & \nodata & \nodata & \nodata & 1.39$_{-0.05}^{+0.05}$ & \nodata & 93.9/129 &  \\
109 & M 59 & PL & $     -5.15$ & 39.30 & \nodata & \nodata & 0.06 & 1.80 & \nodata & 1.4/4 &  \\
110 & M 60 & PL+ME & $     -6.04$ & 38.57 & $     -4.58$ & 39.38 & \nodata & 1.40$_{-2.45}^{+0.76}$ & 1.08$_{-0.04}^{+0.04}$ & 69.4/61 &  \\
111 & M 61 & PL+ME & $     -5.25$ & 39.10 & $     -5.60$ & 38.27 & \nodata & 1.53$_{-0.71}^{+0.46}$ & 0.68$_{-0.24}^{+0.17}$ & 6.3/6 &  \\
112 & M 63 & PL & $     -4.86$ & 38.92 & \nodata & \nodata & 0.17$_{-0.15}^{+0.19}$ & 1.96$_{-0.44}^{+0.52}$ & \nodata & 6.3/9 &
\enddata
\tablecomments{
        Column 1: Source ID.
        Column 2: Galaxy names.
        Column 3: Models, see text for details. 
        Here ``PL'', ``ME'', ``VME'', ``PABS'', ``BB'', ``G'' denote {\tt powerlaw}, 
        {\tt mekal}, {\tt vmekal}, {\tt pcfabs}, {\tt bbody}, 
        and {\tt gaussian} (for iron K$\alpha$ line) component in {\tt XSPEC}, respectively. 
        Column 4: Normalization of {\tt powerlaw} component in fitting model in units of photons keV$^{-1}$ cm$^{-2}$ s$^{-1}$. 
        Column 5: Log of 0.3-10 keV luminosity of {\tt powerlaw} component in fitting model. 
        Column 6: Normalization of {\tt mekal} or {\tt bbody} component in fitting model, if available. It is equal to $(10^{-14}/4\pi D^2)\int n_{\rm e}n_{\rm H} {\rm d}V$ for a {\tt mekal} component, where $D$ is the distance to the source, and $n_{\rm e}$, $n_{\rm H}$ are the electron and hydrogen densities respectively. It is equal to $L_{39}/D_{10}^2$ for a {\tt bbody} component, where $L_{39}$ is the source luminosity in units of $10^{39}$ erg s$^{-1}$, and $D_{10}$ is the distance to the source in units of 10 kpc.
        Column 7: Log of 0.3-10 keV luminosity of {\tt mekal} or {\tt bbody} component in fitting model, if available. 
        Column 8: Intrinsic absorption column density.
        Column 9: Power-law photon index. In a few cases, the errors are not available due to poor statistics.
        Column 10: Temperature of the {\tt mekal} or {\tt bbody} component, if available. 
        Column 11: $\chi^2$ and the degree of freedom.
        Column 12: Note on spectral fitting or reference for pileup sources:
        Ref.1: \citet{ebrero11}, Ref.2: \citet{rosa12}, 
        Ref.3: \citet{cappi06}, Ref.4: \citet{shu10}, 
        Ref.5: \citet{iyomoto98}, Ref.6: \citet{ptak04}, 
        Ref.7: \citet{weaver99}, Ref.8: \citet{guainazzi10}, 
        Ref.9: \citet{evans04}. 
        Table \ref{tab:fit_results} is published in its entirety in the machine-readable format. 
        A portion is shown here for guidance regarding its form and content. 
        } 
\end{deluxetable*}


\boldface{The spectra of two sources (M~64 and M~100) can be well fitted with a single {\tt mekal} component; no power-law component is needed. Figure~\ref{fig:radialprofile} shows their radial profiles, compared with PSFs at the nuclear position, simulated with MARX, and radial profiles of nearby point-like sources. They both appear to be spatially extended. We also examined their lightcurves and did not find any significant flux variability intra or inter observations (significance below $2\sigma$). Thus, these two objects are not considered to be AGN candidates. }

\tablefontsize{\footnotesize}
\begin{deluxetable*}{lrrclrrclrr}[tb]
\tablewidth{0pc} \renewcommand{\arraystretch}{1.2}
\tablecaption{Objects with iron K$\alpha$ line \label{tab:iron}}
\tablehead{\colhead{Name} & \colhead{EW} & \colhead{Ref} & & \colhead{Name } & \colhead{EW } & \colhead{Ref } & & \colhead{ Name} & \colhead{ EW} & \colhead{ Ref}\\ \colhead{ } & \colhead{(eV)} && \colhead{ } & \colhead{ } & \colhead{(eV)}& & \colhead{ } & \colhead{ } & \colhead{(eV)} & \colhead{ }}
\startdata
Circinus Galaxy & 2250 & Ref.2 & & NGC 1365 & 170  &       & & NGC 4939 & 232  &       \\
ESO 428- G 14   & 3041 &       & & NGC 1386 & 2407 &       & & NGC 4945 & 1140 &       \\
IC 2560         & 3319 &       & & NGC 2110 & 88   &       & & NGC 5128 & 82   & Ref.9 \\
MCG -03-34-064  & 636  &       & & NGC 2782 & 629  &       & & NGC 5347 & 521  &       \\
M 51a           & 4793 &       & & NGC 2992 & 255  & Ref.4 & & NGC 5506 & 130  & Ref.8 \\
M 77            & 1200 & Ref.3 & & NGC 3079 & 95   &       & & NGC 5728 & 886  &       \\
M 81            & 40   & Ref.3 & & NGC 4051 & 240  & Ref.3 & & NGC 6300 & 140  & Ref.2 \\
NGC 253         & 851  &       & & NGC 4151 & 300  & Ref.3 & & NGC 7172 & 76   & Ref.2 \\
NGC 1052        & 300  & Ref.7 & & NGC 4388 & 434  &       & & NGC 7479 & 35   &
\enddata
\tablecomments{Reference codes are in Table~\ref{tab:fit_results}.}
\end{deluxetable*}

\section{X-ray Properties of the AGN Candidates}
\label{sec:properties}
\subsection{Luminosity Functions}

The cumulative luminosity functions \boldface{(2$-$10 keV band)} for the AGN candidates are shown in Figure~\ref{fig:lum}, for the whole sample and separately for the different optical spectral classes. The luminosity functions for the whole sample and for Seyferts and LINERs seem to follow a broken power-law form. The small numbers of transition objects, \HII\ nuclei, and absorption-line nuclei preclude a robust determination of their luminosity functions, but they are not inconsistent with a simple power-law. 

\subsection{Photon Index Distribution}
\label{subsec:gamma}

Figure~\ref{fig:gamma} plots the distribution of the photon index $\Gamma$ for objects whose best-fit spectra include a power-law component \boldface{with a degree of freedom (dof) greater than 5}.  Nearly \boldface{3/5 ($\sim$63\%)} of the objects have a $\Gamma$ in the range of 1.3$-$2.3. \boldface{The median value and standard deviation of $\Gamma$ are $1.80\pm0.52$.} \boldface{The standard deviation of the $\Gamma$ distribution in this work is consistent with that in \citet[][$\langle\Gamma\rangle=2.11\pm0.52$]{gonzalez-martin09}, but slightly larger than that in \citet[][$\langle\Gamma\rangle=1.73\pm0.45$]{bianchi09}. We note that the sample in \citet{bianchi09} are all unobscured AGNs.  If we exclude those objects with \nhint\ $>0.5\times10^{22}$~cm$^{-2}$ in our sample, the standard deviation of $\Gamma$ becomes 0.46, consistent with the results of \citet{bianchi09}.  The discussion for $\Gamma$ and AGN accretion states will be detailed in a forthcoming paper (She et al.\ in preparation). } 

\section{Contamination by X-ray Binaries}
\label{sec:contamination}

The AGN candidates, especially the low-luminosity sources, identified via near-infrared/optical and X-ray cross-correlation may be contaminated by other types of X-ray sources that are coincident with the galaxy nucleus. The most likely contaminants to LLAGNs are X-ray binaries. It is difficult to individually distinguish an LLAGN from a nuclear X-ray binary without deep and/or long-term observations.  Nevertheless, we argue that statistically the vast majority of these nuclear point-like X-ray sources are indeed AGNs.

As shown in Figure~\ref{fig:offsets}, the distribution of positional offsets between the stellar nucleus and the nearest X-ray source on the sky for all galaxies in our sample is bimodal.  Our identified AGN candidates mostly occupy the first peak, consistent with little to no positional offset. This indicates that the nuclear X-ray  sources belong to a population physically associated with the galaxy nuclei, ruling out the possibility  that they belong to a distribution extended throughout the host galaxy that have been identified due to random coincidence with the nucleus. There are \onesec\ AGN candidates located on the sky plane less than 1\arcsec\ to the near-infrared/optical nuclei. For comparison, the number density for point-like X-ray sources between 5\arcsec\ to 10\arcsec\ around the near-infrared/optical nuclear position is $\sim$60 times lower in our sample galaxies.  Of course, we cannot definitively rule out that our AGN candidates are accreting X-ray binaries residing within nuclear star clusters, a common component seen in late-type galaxies (e.g., \citealt{boker02}). For example, \citet{feng15} discovered a luminous X-ray flare from the nucleus of NGC 247, which could be due to an outburst from a low-mass X-ray binary in the nuclear star cluster.

The luminosity functions of the nuclear X-ray sources (Figure~\ref{fig:lum}) also suggest that most of the candidates are AGNs. For comparison, the slope of the luminosity function is about $-0.61$ for high-mass X-ray binaries \citep{grimm03} and around $-1.1$ for low-mass X-ray binaries \citep{kim04}. For ultraluminous X-ray sources, the slope is around $-0.5$ to $-0.8$, depending on the choice of model (power-law or cutoff power-law) or the luminosity range \citep{swartz11}, because their luminosity function rolls over and cuts off around $10^{41}$~\ergs.  The luminosity function for our combined sample of all nuclear X-ray sources is much shallower than that of X-ray binaries, over the luminosity range in which they overlap.  The luminosity function for the AGN candidates without line emission (absorption-line nuclei) do appear similar to that of ultraluminous X-ray sources.  The nature of these sources will be discussed in depth in forthcoming papers. 

\subsection{Stacking Non-detections}

To evaluate whether the vast majority of the AGN candidates are indeed AGNs, we performed a stacking analysis of the images for which no significant nuclear X-ray emission was detected in the individual exposures. For the non-detections, we stacked their exposure-corrected flux images centered at the near-infrared/optical position of each galaxy, after excising all nearby point-like sources detected by \textit{wavdetect}. We excluded those images on which a point-like source was detected within 5$\arcsec$ radius from the galaxy nucleus; these images \boldface{may contaminate the central region}.  Our final stack contains images of 368 non-detections, with a total exposure time of 6935.2 ks.  The stacked 0.3$-$8 keV flux image, shown in Figure~\ref{fig:stackimage}, reveals a statistically significant source with a net photon flux of $5.1\times10^{-7}$ \phcms\, \boldface{from} an aperture of 2$\arcsec$ radius and background from an annulus with inner and outer radii of 15$\arcsec$ and 25$\arcsec$, respectively. This corresponds to a flux of $8.9\times 10^{-16}$ \ergcms, given a power-law model with a photon index $\Gamma=1.8$ and a typical Galactic absorption \nhgal=$2\times10^{20}$~cm$^{-2}$. \boldface{The aperture size was chosen in view of the fact that the typical PSF radius of {\it Chandra}/ACIS is less than 1$\arcsec$, the absolute astrometry is better than 1\arcsec, and most of the source extracting regions given by {\it wavdetect} have a radius less than 2$\arcsec$. The flux will be 3 times higher if an aperture with a radius of 5$\arcsec$ is used.} \boldface{To evaluate if observations with different exposures and backgrounds will bias the sensitivity,  we split all non-detections into several subsets depending on their exposure times and performed the same stacking analysis. The derived flux from different subsets follows $F \propto t^{-1/2}$, where $t$ is total exposure of the subset, suggesting that there is no significant bias from stacking both shallower and deeper exposures.} We checked the stacked images in two bands, 0.3$-$2 and 2$-$8 keV, and found that the fluxes of the X-ray excess in these two bands are consistent with emission from such an absorbed power-law model, but inconsistent with emission from hot gas ({\tt mekal} with a temperature of 0.4$-$0.9 keV). At the typical distance of 17 Mpc for the 368 galaxies with non-detections, this yields an average X-ray luminosity of $3.1\times 10^{37}$ \ergs.  This luminosity is broadly consistent with that of an X-ray binary, and thus this emission can be accounted for by an ensemble of low-luminosity X-ray binaries. The spatial extension of the X-ray excess ($R \sim 5\arcsec$) is also larger than the concentration of the detected AGNs (mostly within 1\arcsec), suggesting that they are likely X-ray sources associated with the nuclear stellar component. 

The above results further strengthen our argument that the X-ray-detected nuclear sources do not suffer significantly from contamination by X-ray binaries or emission from hot gas. The above calculated average luminosity contributes only $\sim$1\% of a typical AGN candidate luminosity (0.3$-$8 keV; $2.5\times 10^{39}$ \ergs) in this work.

\acknowledgments
We thank the anonymous referee for his/her insightful suggestions to improve this paper. The work of LCH is supported by National Key Program for Science and Technology Research and Development grant 2016YFA0400702. HF acknowledges funding support from the National Natural Science Foundation of China under grant No.\ 11633003, and the Tsinghua University Initiative Scientific Research Program. This research has made use of the NASA/IPAC Extragalactic Database (NED) which is operated by the Jet Propulsion Laboratory, California Institute of Technology, under contract with the National Aeronautics and Space Administration. We also acknowledge the usage of the HyperLeda database (http://leda.univ-lyon1.fr).


\end{document}